\documentclass[prb,aps,twocolumn,floatfix,showpacs,longbibliography]{revtex4-1}
\usepackage{amssymb}
\usepackage{amsmath}
\usepackage{graphicx}
\usepackage[colorlinks=true,linkcolor=blue,anchorcolor=red,citecolor=blue,urlcolor=blue]{hyperref}

\begin{document}

\title{Edge states and integer quantum Hall effect in topological insulator
thin films}

\author{Song-Bo Zhang, Hai-Zhou Lu, \& Shun-Qing Shen}

\affiliation{Department of Physics, The University of Hong Kong, Pokfulam Road,
Hong Kong, China}

\date{\today }
\begin{abstract}
The integer quantum Hall effect is a topological state of quantum
matter in two dimensions, and has recently been observed in three-dimensional
topological insulator thin films. Here we study the Landau levels
and edge states of surface Dirac fermions in topological insulators
under strong magnetic field. We examine the formation of the quantum
plateaux of the Hall conductance and find two different patterns,
in one pattern the filling number covers all integers while only odd
integers in the other. We focus on the quantum plateau closest to
zero energy and demonstrate the breakdown of the quantum spin Hall
effect resulting from structure inversion asymmetry. The phase diagrams
of the quantum Hall states are presented as functions of magnetic
field, gate voltage and chemical potential. This work establishes
an intuitive picture of the edge states to understand the integer
quantum Hall effect for Dirac electrons in topological insulator thin
films.
\end{abstract}
\maketitle
\ \ \\
 \textbf{Introduction}\\
 The discovery of the integer quantum Hall effect in two-dimensional
electron gas opens a window to explore topological phases in quantum
matters\cite{Klitzing80prl,QHE-book}. In the quantum Hall effect
the longitudinal conductance vanishes while the Hall conductance $\sigma_{xy}$
is quantized at $\nu e^{2}/h$, where $e$ is the elementary charge
and $h$ is the Planck constant. It is known that the integer $\nu$
is the topological invariant of a quantum phase, it counts the number
of conducting chiral channels at the edges of the system, and is insensitive
to the geometry of the sample, impurity, and interactions of electrons
\cite{TNKK_1982prl,Halperin82prb,MacDonald84prb,Hatsugai_1993prl}.
Three-dimensional topological insulator is a new class of topological
material and is characterized by the formation of the Dirac fermion
gas covering its surfaces \cite{moore2010Nature,Hasan10rmp,QiXL2011rmp,Shen_book}.
Soon after the discovery of three-dimensional topological insulators,
the formation of the Landau level (LL) of the surface Dirac electrons
in a strong magnetic field has been observed by the scanning tunneling
microscope \cite{QuDongxia2010science,Analytis2010NatPhys,Hanaguri2010prb,Chengpeng2010prl,FuYishuang_2014nphys}
and Shubnikov-de Haas oscillations in the longitudinal conductance
\cite{QuDongxia2010science,Analytis2010NatPhys,Brune2011prl}. It
has been known since 1980's that the Hall conductance of massless
Dirac fermions is quantized as half integers $\nu=n+1/2$, where $n=0,\pm1,\pm2,...$.
\cite{Jackiw84prd,Schakel91prd,Ando98jpsj,Gusynin_2005prl,LeeDH_2009prl},
and $1/2$ is attributed to the Berry phase of the massless Dirac
fermions acquiring from a cyclotron motion around the Fermi surface
\cite{Ando98jpsj,Mikitik1999prl,Igor2004prl}. Usually each LL carries
one conducting channel near the edge due to the geometric distortion
of the cyclotron motion of electrons in a magnetic field. The quantum
Hall conductance is determined by the number of the edge states \cite{Halperin82prb,MacDonald84prb}.
Thus the relation between the half-quantized Hall conductance and
the number of the conducting edge channel is an open issue and has
attracted a lot of studies \cite{Gusynin_2005prl,Dmitry2006prl,Peres2006prb,Zhoubin2006prb,QiXL2008prb,QiXL2009science,LeeDH_2009prl,Tkachov2010prl,Chu-11prb,Vafek2011prb,Zyuzin2011prb,Sitte2012prl,ZhangYY2012JPCM,liGuohong2013natcomm,Konig2014prb}.
Very recently the integer quantum Hall effect has been measured in
three-dimensional topological insulator thin films by two independent
groups \cite{XuYang2014NatPhys,Yoshimi_2014arXiv}. One group measured
a series of plateaux of $\nu=-1,0,$1,2,3 and $\nu=1,3$ in $\mathrm{BiSbTeSe}_{2}$
\cite{XuYang2014NatPhys}, and the other measured the plateaux of
$\nu=0,\pm1$ in $\left(\mathrm{Bi}_{1-x}\mathrm{Sb}_{x}\right)_{2}\mathrm{Te_{3}}$
\cite{Yoshimi_2014arXiv}. These plateaux of the Hall conductance
are attributed to the addition of the top and bottom surface electrons,
and always yield integers in units of $e^{2}/h$.

In this work we present solutions of the LLs and edge states of the
surface electrons, and explore the formation of the quantum Hall effect
in a topological insulator thin film. The Hall conductance is calculated
by means of the Kubo formula at zero temperature, which is well quantized
when the chemical potential lies between two LLs. Two distinct patterns
of the quantum Hall conductance are found: $\nu$ is an odd integer
when the LLs of the top and bottom surface electrons are degenerate
in a thick film, and $\nu$ is an integer when structure inversion
symmetry is broken in the film or the top and bottom surface electrons
are coupled to open an energy gap due to the finite-size effect \cite{Lu10prb,Shan11njp}.
The absence of the $\nu=0$ plateau is caused by the degeneracy of
two sets of LLs, and the $\nu=0$ plateau emerges when the degeneracy
is lifted by the finite-size effect or structure inversion asymmetry
(SIA, which can be produced by the energy difference between the two
surfaces). The width of the $\nu=0$ plateau is determined by the
energy difference between the two LLs closest to zero energy. The
state of $\nu=0$ can be either in the quantum spin Hall phase or
trivial band insulator phase, and the quantum spin Hall effect can
be easily broken down by SIA.

\ \ \\
 \textbf{Results}\\
 \textbf{Model.} Consider a thin film of three-dimensional topological
insulator. Starting from the bulk Hamiltonian for topological insulators
\cite{ZhangHJ_09nphys,Shen_book}, the low-energy effective Hamiltonian
for the surface electrons has been derived by solving the differential
equation for the bulk bands exactly \cite{Lu10prb,Shan11njp}
\begin{equation}
H=\begin{pmatrix}\frac{\Delta}{2}-Bk^{2} & i\gamma k_{-} & V & 0\\
-i\gamma k_{+} & -\frac{\Delta}{2}+Bk^{2} & 0 & V\\
V & 0 & -\frac{\Delta}{2}+Bk^{2} & i\gamma k_{-}\\
0 & V & -i\gamma k_{+} & \frac{\Delta}{2}-Bk^{2}
\end{pmatrix},\label{Hamiltonian}
\end{equation}
where $k_{\pm}=k_{x}\pm ik_{y}$ and $k^{2}=k_{x}^{2}+k_{y}^{2}$
with $k_{x,y}$ the wave vector in the surface plane. The mass term
$\Delta/2-Bk^{2}$ is generated by the hybridization between the wave
functions of the top and bottom surface electrons. Both $B$ and $\Delta$
depend on the film thickness and approach zero simultaneously in a
thick film. $\gamma=v_{F}\hbar$ with $v_{F}$ the effective velocity
and $\hbar$ the reduced Planck constant. $2V$ is the potential difference
between the top and bottom surfaces due to SIA as shown in Fig. \ref{Fig1:2Patterns}a,
which can be induced in a realistic thin film by the potential difference
between the substrate and vacuum surfaces, and tunable by a gate voltage.
The interplay of SIA and top-bottom hybridization gives the Rashba-like
splitting in the band structure (see Fig. \ref{Fig1:2Patterns}b).
The physics of this model has been confirmed by the angle-resolved
photoemission spectroscopy experiments on topological insulator thin
films \cite{ZhangY_10nphys,Sakamoto10prb}.

\ \ \\
 \textbf{Two patterns of quantum Hall plateaux}. For a thin film of
topological insulator in the presence of a perpendicular magnetic
field, the Hall conductance is usually quantized as an integer $\nu$
$\left(=0,\pm1,\pm2,\cdots\right)$ as shown in Fig. \ref{Fig1:2Patterns}c,
or can only be an odd integer $\nu$ $\left(=\pm1,\pm3,\cdots\right)$
when the LLs of the top and bottom surface electrons coincide exactly
as shown in Fig. \ref{Fig1:2Patterns}d. Consider a topological insulator
thin film. The surface states cover the top and bottom surfaces as
illustrated in Figs. \ref{Fig1:2Patterns}a and \ref{Fig1:2Patterns}b.
The lateral side is also covered by the surface electrons, and is
ignored in the present work. For a film with a relatively large thickness,
the top and bottom surface states are well separated. Thus the system
consists of two decoupled massless Dirac electrons. The two Dirac
points can be separated by the potential difference $2V$ between
the top and bottom surfaces. The value of $V$ then can determine
the quantized pattern of the Hall conductance such as the width of
quantum plateaux and the existence of $\nu=0$ plateau. For $V=0$,
the LLs of the top and bottom surface electrons are degenerate, and
the Hall conductance is quantized as odd integers $\nu=2n+1$, which
can be regarded as the addition of two sets of half-quantized Hall
conductance, i.e., $\nu=2(n+1/2)$. This is very similar to that in
graphene, in which the Hall conductance is $\nu=4(n+1/2)$, where
the factor 4 is attributed to the spin and valley degrees of freedom
\cite{Novoselov2005Nature,ZhangYB2005Nature}. For $V\neq0$, the
degeneracy of the LLs is removed, and the Hall conductance becomes
quantized as integers in Fig. \ref{Fig1:2Patterns}c. Another factor
leading to the lift of the degeneracy is the finite-size effect. When
the thickness of a thin film is comparable with the spatial distribution
of the wave functions of the surface states, the overlap of the wave
functions will open an energy gap $\Delta$ at the Dirac points of
the two surface states \cite{Lu10prb}, leading to the presence of
the $\nu=0$ plateau and a possible topological phase of the quantum
spin Hall effect.

\ \ \\
 \textbf{Landau levels and edge states}. In Figs. \ref{Fig2:LL-noSIA}
and \ref{Fig3:LL-SIA}, we present the LLs in a magnetic field $\mu_{0}H$
normal to the thin film, the LL energies or edge states near one edge
of the system, and the corresponding patterns of the quantum Hall
conductance. In the absence of SIA, i.e., $V=0$, four possible typical
cases are shown in Fig. \ref{Fig2:LL-noSIA}. Case (i) is for a thick
film, i.e., $\Delta=B=0$. In the bulk, the LLs of zero energy are
degenerate, and is insensitive to the field, as shown in Fig. \ref{Fig2:LL-noSIA}a.
However they split into two branches when approaching one edge: one
branch goes upward (called electron-like) and the other goes downward
(called hole-like), as shown in Fig. \ref{Fig2:LL-noSIA}e. The position
$y_{0}=\ell_{B}^{2}k_{x}$ is the guiding center of the wave packages
of surface LLs and is proportional to the wave vector $k_{x}$, where
the magnetic length $\ell_{B}=\sqrt{\hbar/e\mu_{0}H}$. The slope
of the energy dispersion near the edge $\partial E/\partial y_{0}=(1/\ell_{B}^{2})(\partial E/\partial k_{x})$
is proportional to the effective velocity of the edge states $v_{\text{{eff}}}=(1/\hbar)(\partial E/\partial k_{x})$,
which defines the current flow of the edge states. The Hall conductance
is equal to $\nu=1$ when the chemical potential is below the LLs
of $n=0$ and $\nu=-1$ when the chemical potential is above the LLs
of $n=0$. The plateau of $\nu=0$ is absent and other plateaux are
$\nu=\pm3,\pm5,\cdots$. For a thinner film, there exist three cases,
(ii)$\Delta>0$, (iii) $\Delta=0$, and (iv) $\Delta<0$, all with
the parameter $B<0$ (without loss of generality we assume negative
$B$). In case (ii) with $\Delta B<0$, the two LLs of $n=0$ are
separated in a finite field as shown in Fig. \ref{Fig2:LL-noSIA}b.
When approaching the edge the LLs with positive energies go upward
while the LLs with negative energies go downward, indicating that
the edge electrons with opposite energies move in opposite directions.
One of the key features of the quantum Hall conductance is the emergence
of the $\nu=0$ plateau, and the width of the plateau is determined
by the energy difference between the two LLs of $n=0$ (see Fig. \ref{Fig2:LL-noSIA}j).
In case (iii) with $\Delta=0$ and $B<0$, the two LLs near the Dirac
point are degenerate at zero field, as shown in Fig. \ref{Fig2:LL-noSIA}c.
The degeneracy is lifted by a finite field $\mu_{0}H$. The energies
of edge states (see Fig. \ref{Fig2:LL-noSIA}g) and the Hall conductance
(see Fig. \ref{Fig2:LL-noSIA}k) are very similar to those in case
(ii). In case (iv) with $\Delta B>0$, two LLs near zero energy are
separated in weak fields, and cross at a finite field as shown in
Fig. \ref{Fig2:LL-noSIA}d, indicating a quantum phase transition.
In a weak field, the pattern of the Hall conductance (see Fig. \ref{Fig2:LL-noSIA}l)
is also similar to that of case (ii), while the dispersions of the
edge states (see Fig. \ref{Fig2:LL-noSIA}h) are different. The two
LLs of $n=0$ cross near the edge, which is a key feature of the quantum
spin Hall effect in a finite field \cite{Konig2007science}. At magnetic
fields higher than the energy crossing in Fig. \ref{Fig2:LL-noSIA}d,
the LLs near zero energy never cross near the edge, similar to those
in Figs. \ref{Fig2:LL-noSIA} f and g, indicating the breakdown of
the quantum spin Hall effect in a magnetic field. However, the plateau
of $\nu=0$ still survives.

In the presence of SIA, $V\neq0$, the relative positions of two sets
of LLs from the top and bottom surfaces can be tuned by the value
of $V$. A typical pattern of the quantum Hall conductance is presented
in Fig. \ref{Fig3:LL-SIA}. The Hall conductance covers all integers.
In the case that the chemical potential crosses some accidental crossing
points of two LLs, the Hall conductance can change by $2e^{2}/h$.

\ \ \\
 \textbf{Quantum spin Hall state and its breakdown}. A question arises
from the appearance of the quantum spin Hall effect in case (iv) with
$\Delta B>0$. Usually the quantum spin Hall effect is protected by
time reversal symmetry \cite{Kane2005prl,Bernevig2006science}. Applying
a magnetic field breaks time reversal symmetry, but it is known from
the calculation of the Hall conductance that the quantum spin Hall
effect can be stabilized in a finite field when the system possesses
an intrinsic and hidden $\mathrm{s}_{z}$ symmetry which guarantees
the decoupling of the two blocks with spin up and down in the Hamiltonian
\cite{ZhangSB14prb}. In the presence of SIA, $V\neq0$, the situation
changes. The energy crossing of the two $n=0$ LLs in a finite field
in Fig. \ref{Fig2:LL-noSIA}d becomes an anti-crossing for a finite
$V$ in Fig. \ref{Fig4:BreakDown}a, showing that there is no field-induced
quantum phase transition from the quantum spin Hall insulator to the
trivial band insulator in the presence of SIA and the system remains
in the trivial phase at all magnetic fields. In other words, the quantum
spin Hall effect breaks down as long as SIA is present. The breakdown
of the quantum spin Hall phase can also be seen from the energies
of the edge states in Fig. \ref{Fig4:BreakDown}b, in which the energies
of the $n=0$ LLs cross for $V=0$ while open a finite gap for a nonzero
$V$. The gap increases with increasing $V$. For a small $V$, the
corresponding energy gap of the edge states is also small. Although
the edge states are no longer protected topologically, it can still
produce some physical phenomena such as the spin accumulation, or
the intrinsic spin Hall effect near the edge when an electric current
is applied. However these effects will be weakened for a large $V$.
Therefore, the quantum spin Hall effect does not survive in the presence
of both a magnetic field and SIA.

\ \\
 \textbf{Phase diagrams}. In the absence of SIA, i.e., $V=0$, we
plot in Figs. \ref{Fig5:PhaseDiagrams}a-d the corresponding phases
diagrams of the Hall conductance as a function of the chemical potential
$\mu$ and magnetic field $\mu_{0}H$ for the four cases. Different
quantum Hall phases are denoted by the quantized Hall conductances
and are separated by the boundaries (marked by the white dotted lines).
The Hall conductance is antisymmetric with respect to $\mu$. In case
(i), there are only odd integer quantum Hall phases (see Fig. \ref{Fig5:PhaseDiagrams}a).
The spacings of the phases grow with increasing magnetic field or
$|\mu|$. Thus the quantum Hall plateaux are more visible by varying
$\mu$ in a stronger magnetic field, or by varying the magnetic field
with a larger fixed $|\mu|$. In contrast, not only odd but also even
integer quantum Hall phases are possible in the presence of a finite
mass, i.e., $B\neq0$ or $\Delta\neq0$ (see Figs. \ref{Fig5:PhaseDiagrams}b-d).
When $\mu=0$, all negative LLs are filled and all positive LLs are
empty no matter how large the magnetic field is, and the corresponding
Hall conductance is zero, regardless of whether there is energy crossing
near the edge of the system. The quantum spin Hall effect only appears
in the case of $\Delta B>0$ and $V=0$ in a weak field, and disappears
in a stronger field as shown in Fig. \ref{Fig5:PhaseDiagrams}d.

Moreover, we calculate the Hall conductance as a function of $V$
and magnetic field $\mu_{0}H$ in Figs. \ref{Fig5:PhaseDiagrams}e-h,
where we set $\mu=-V-0^{+}$. An infinitesimal value $0^{+}$ is introduced
to avoid the alignment of the chemical potential $\mu$ with one of
the $n=0$ LLs $(E_{0}=-V)$ in case (i) in Fig. \ref{Fig5:PhaseDiagrams}e.
The asymmetry of the Hall conductance with respect to $V$ reflects
the fact that both of the $n=0$ LLs $\left(E_{0}=\pm V\right)$ are
empty for a positive $V$ while only one level $\left(E_{0}=+V\right)$
is filled for a negative $V$. Figs. \ref{Fig5:PhaseDiagrams}f-h
are for the cases with a finite mass, the Hall conductance is antisymmetric
with respect to $V$ and show very similar phase diagrams. Both odd
and even integer quantum Hall phases can be induced by changing $V$
in all the four cases. If setting $\mu=0$, we find that the Hall
conductance is always vanishing, which is expected since the chemical
potential $\mu$ is fixed at the center between the electron-like
LLs that would go upward and the hole-like LLs that go downward when
approaching the edge.

\ \\
 \textbf{Discussions}\\
 The formation of the edge states related to two LLs near zero energy
can be understood from the model that the top and bottom surfaces
are separated by the metallic lateral surface. In the presence of
a perpendicular magnetic field, the lateral surface electrons only
experience an in-plane field, and the two Dirac points will shift
oppositely by a constant if the Zeeman field is taken into account
\cite{Chu-11prb}. The edge states for the two LLs near the zero energy
are divided into three parts: the parts of the wave function at the
top and bottom surfaces decay exponentially away from the edge while
they are connected by the part of the lateral surface electron. This
is very similar to the case of the quantum anomalous Hall effect in
a topological insulator thin film with a perpendicular Zeeman field\cite{Chu-11prb}.
A similar calculation can be found in a recent paper \cite{Morimoto2014xxx}.
As the conductance of lateral surfaces is nonzero for a thin film
with a finite thickness, the longitudinal conductance no longer vanishes.
In this case the Hall resistance can be quantized perfectly only when
the residual conductance from the lateral effect can be suppressed
completely.

Mathematically the coupling between the top and bottom surfaces has
been reasonably taken into account in the model in Eq. (\ref{Hamiltonian}).
Although it was shown rigorously that the Hall conductance for an
ideal massless Dirac fermion gas with $H_{D}=v_{F}(k_{x}\sigma_{y}-k_{y}\sigma_{x})$
is quantized in a magnetic field as half integers, it is found that
the Hall conductance is modified into integers once a quadratic correction
$\triangle H=-B(k_{x}^{2}+k_{y}^{2})\sigma_{z}$ is introduced to
$H_{D}$, where $\sigma_{x,y,z}$ are the Pauli matrices. This is
case (iii) in Fig. \ref{Fig2:LL-noSIA}c. The conclusion is valid
even if the parameter $B$ is in the infinitesimal limit, and the
solution of the edge states always exists. In other words, for an
infinitesimal $B,$ we obtain two well-defined edge states mathematically
for two LLs of $n=0$ as shown in Fig. \ref{Fig2:LL-noSIA}e. The
physical meaning of the term $B$ is attributed to the coupling between
the top and bottom surfaces according to the exact solution for a
bulk model \cite{Lu10prb,Shan11njp}.

Having in mind the picture of edge states of LLs for Dirac fermions
we come to make some comments on the two experiments on the quantum
Hall effect in topological insulator thin films. In the experiment
by Xu et al\cite{XuYang2014NatPhys}, a series of the Hall conductance
plateaux of $\nu=-1,0,1,2,3$ are measured as a function of the gate
voltage $V_{bg}$. The absence of the $\nu=-2,-3$ plateaux can be
understood as the particle-hole symmetry breaking in the band structure
of $\mathrm{BiSbTeSe}_{2}$, which is also confirmed by the transport
measurement. The role of $V_{bg}$ is to control the relative positions
of the two Dirac points at the top and bottom surfaces, or the SIA
term $V$ in the thin film model in Eq. (\ref{Hamiltonian}). The
nonzero conductance $\sigma_{xx}$ indicates that the Hall resistance
has not yet been quantized completely. One reason is that the thicknesses
of the samples are 80 nm and 160 nm, respectively. In this case the
lateral conductance does not vanish completely even at low temperatures.
Suppression of the lateral effect will be a key to realize high precision
of quantum Hall conductance in this experiment. On the odd integer
quantum Hall effect, the plateaux of $\nu=1$ and 3 are observed.
As one surface of the thin film is grown on the substrate while the
other is exposed to the vacuum, the boundary conditions are quite
different, and the effective velocities of the surface electrons may
not be identical. Thus it will be a hard task to make two sets of
the LLs of the surface electrons degenerate completely by tuning the
gate voltage only. However it is relatively easy to have only two
specific LLs degenerate such that the plateau of $\nu=0$ disappears.

In the experiment by Yoshimi et al\cite{Yoshimi_2014arXiv}, only
two plateaux of $\nu=-1,+1$ or $\nu=0,+1$ are measured in two different
samples of $\left(\mathrm{Bi}_{1-x}\mathrm{Sb}_{x}\right)_{2}\mathrm{Te_{3}}$.
The presence of $\nu=0$ plateau is attributed to the potential difference
between the two surfaces, i.e., the SIA term $V\neq0$. As the thickness
of two samples is 8 nm, the finite-size effect is relatively weak,
i.e., $\Delta-Bk^{2}\rightarrow0$. This should correspond to the
case in Fig. \ref{Fig3:LL-SIA}. Thus the width of the plateau is
determined by the value of $V$, and the pattern of edge states indicates
that the state of $\nu=0$ is simply a band insulator.

\ \\
 \textbf{Methods}\\
 \begin{small} \textbf{Landau levels.} When a uniform field is applied
perpendicular to the thin film, the wave vector is replaced by $\mathbf{k}\rightarrow-i\nabla+e\mathbf{A}/\hbar$,
where the vector potential under the Landau gauge is $\mathbf{A}=(-\mu_{0}Hy,0,0)$.
$k_{x}$ remains a good quantum number. The LLs can be found by defining
two ladder operators $a=-(y/\ell_{B}+\partial_{y}\ell_{B}-\ell_{B}k_{x})/\sqrt{2}$
and $a^{\dag}=-(y/\ell_{B}-\partial_{y}\ell_{B}-\ell_{B}k_{x})/\sqrt{2}$,
where the magnetic length $\ell_{B}=\sqrt{\hbar/e\mu_{0}H}$, and
assuming the trial solution,
\begin{eqnarray}
\psi_{n}(x,y) & = & \dfrac{1}{\sqrt{L_{x}}}e^{ik_{x}x}\begin{pmatrix}C_{n1}\langle y|n-1\rangle\\
C_{n2}\langle y|n\rangle\\
C_{n3}\langle y|n-1\rangle\\
C_{n4}\langle y|n\rangle
\end{pmatrix},~~n\geqslant1;\label{eq:1}\\
\psi_{0}(x,y) & = & \dfrac{1}{\sqrt{L_{x}}}e^{ik_{x}x}\begin{pmatrix}0\\
C_{01}\langle y|0\rangle\\
0\\
C_{02}\langle y|0\rangle
\end{pmatrix},~~n=0,
\end{eqnarray}
where $\langle y|n\rangle$$=\left(1/\sqrt{n!2^{n}\ell_{B}\sqrt{\pi}}\right)\exp[-(y-y_{0})^{2}/2\ell_{B}^{2}]\mathcal{H}_{n}[(y-y_{0})/\ell_{B}]$
and $\mathcal{H}_{n}$ are Hermite polynomials. The energies of the
LLs are found as
\begin{eqnarray}
E_{n,s}^{\pm} & = & \pm\sqrt{(\epsilon_{n}+s\mathcal{P}_{n})^{2}+V^{2}\cos^{2}\Theta_{n}},\ \ n\geqslant1;\\
E_{0}^{\pm} & = & \pm\sqrt{(-\Delta/2+\omega/2)^{2}+V^{2}},\ \ n=0,
\end{eqnarray}
where $\epsilon_{n}$=$\sqrt{n\eta^{2}+(\Delta/2-n\omega)^{2}}$,
$\mathcal{P}_{n}$=$\sqrt{(\omega/2)^{2}+V^{2}\sin^{2}\Theta_{n}}$,
$\cos\Theta_{n}$=$(n\omega-\Delta/2)/\epsilon_{n}$, $\sin\Theta_{n}$=$\sqrt{n}\eta/\epsilon_{n}$,
$s=\pm1$, $\omega=2B/\ell_{B}^{2}$, and $\eta=\sqrt{2}A/\ell_{B}$.
For each given $n\geqslant1$, there are four LLs, two of them ($E_{n,s=\pm1}^{+}$)
are electron-like while the other two ($E_{n,s=\pm1}^{-}$) are hole-like.
The expressions for the corresponding eigenstates can be found in
the supplementary information.

The magnetic field can also induce a Zeeman energy described by $H_{z}=(g\mu_{B}\mu_{0}H/2)\sigma_{0}\otimes\sigma_{z}$
where $\mu_{B}$ is the Bohr magneton and $g$ is the g-factor. It
is weak, and thus we neglect it in this work.

\ \\
 \textbf{Hall conductance.} The Hall conductance can be found from
the Kubo formula
\begin{eqnarray}
\sigma_{xy}=\dfrac{e^{2}\hbar}{2\pi\ell_{B}^{2}}\sum_{n\alpha s\neq m\beta t} & \mathrm{Im}\big[\langle n\alpha s|\hat{v}_{x}|m\beta t\rangle\langle m\beta t|\hat{v}_{y}|n\alpha s\rangle\big]\nonumber \\
 & \times\dfrac{n_{F}(E_{n,s}^{\alpha}-\mu)-n_{F}(E_{m,t}^{\beta}-\mu)}{(E_{n,s}^{\alpha}-E_{m,t}^{\beta})^{2}},
\end{eqnarray}
where $|n\alpha s\rangle$ is the eigenstates in the LL $E_{n,s}^{\alpha}$,
$n_{F}(E)=1/[1+\exp(E/k_{B}T)]$ is the Fermi function with $k_{B}$
the Boltzmann constant. $\hat{v}_{x}=(-i/\hbar)[x,H]$ and $\hat{v}_{y}=(-i/\hbar)[y,H]$
are two velocity operators.

\ \\
 \textbf{Edge states at open boundary.} To solve the wave function
of edge states at an open boundary, we employ the properties of the
two standard solutions $U_{\lambda}(\xi)$ and $V_{\lambda}(\xi)$
to the Weber equation \cite{Abramowitz_book},
\begin{eqnarray}
 &  & a^{\dag}U_{\lambda}(\xi)=-U_{\lambda-1}(\xi),\label{Eq:recurrence1}\\
 &  & aU_{\lambda}(\xi)=\big(\lambda+1/2\big)U_{\lambda+1}(\xi),\label{Eq:recurrence2}\\
 &  & a^{\dag}V_{\lambda}(\xi)=\big(\lambda-1/2\big)V_{\lambda-1}(\xi),\label{Eq:recurrence3}\\
 &  & aV_{\lambda}(\xi)=-V_{\lambda+1}(\xi),\label{Eq:recurrence4}
\end{eqnarray}
where the dimensionless quantity is defined by $\xi=(y-k_{x}\ell_{B}^{2})\sqrt{2}/\ell_{B}$.
With the trial wave functions in the $y$ direction
\begin{align}
\varphi_{u}(\lambda,\xi)=\begin{pmatrix}u_{1}U_{\lambda}(\xi)\\
u_{2}U_{\lambda-1}(\xi)\\
u_{3}U_{\lambda}(\xi)\\
u_{4}U_{\lambda-1}(\xi)
\end{pmatrix},\ \ \varphi_{v}(\lambda,\xi)=\begin{pmatrix}v_{1}V_{\lambda}(\xi)\\
v_{2}V_{\lambda-1}(\xi)\\
v_{3}V_{\lambda}(\xi)\\
v_{4}V_{\lambda-1}(\xi)
\end{pmatrix},
\end{align}
we can find the corresponding $\lambda$ for a given eigen energy
$E$. There are four $\lambda$'s for each given eigenenergy $E$,
so a general solution to the wave function $\Psi(E,\xi)$ is a linear
combination of eight eigenstates. The allowed eigen energies, the
superposition coefficients as well as the wave function can be found
by applying the open boundary conditions at the edge.

For a system with a semi-infinite geometry $y\in[0,+\infty)$, the
wave function $\Psi(E,\xi)$ can only contain the $U_{\lambda}(\xi)$
components since $V_{\lambda}(\xi)$ is exponentially divergent while
$U_{\lambda}(\xi)$ are vanishing as $\xi$ approaches $+\infty.$
The boundary condition at $y=0$ further provides an equation for
$E$. In the weak coupling limit $\Delta,B\rightarrow0$, the equation
for $E$ is reduced as
\begin{align}
\Big(\dfrac{E}{\eta}\Big)^{2}=\Big(\dfrac{V}{\eta}\Big)^{2}+\dfrac{U_{\lambda_{1}-1}(\xi_{0})}{U_{\lambda_{1}}(\xi_{0})}\dfrac{U_{\lambda_{2}-1}(\xi_{0})}{U_{\lambda_{2}}(\xi_{0})},
\end{align}
where $\lambda_{1,2}=1/2-(E\pm V)^{2}/\eta^{2}$ and $\xi_{0}=-k_{x}\ell_{B}^{2}$.
If $E$ is a solution, then $-E$ is also a solution, reflecting the
particle-hole symmetry, as expected. Thus we shows that the solutions
of edge states exist even if $B\rightarrow0^{\pm}$.

\end{small}


\ \ \\
 \textbf{Acknowledgements}\\
 This work is supported by Research Grants Council, University Grants
Committee, Hong Kong, under Grant No. 17304414.

\ \ \\
 \textbf{Author contributions}\\
 S.B.Z. performed the calculations in the paper with assistance from
H.Z.L and S.Q.S. All authors wrote the paper. S.Q.S planned and supervised
the project.

\ \\
 \textbf{Additional information}\\
 Competing financial interests: The authors declare no competing financial
interests. \begin{widetext} \newpage{}

\begin{figure}[tbph]
\centering \includegraphics[width=0.4\textwidth]{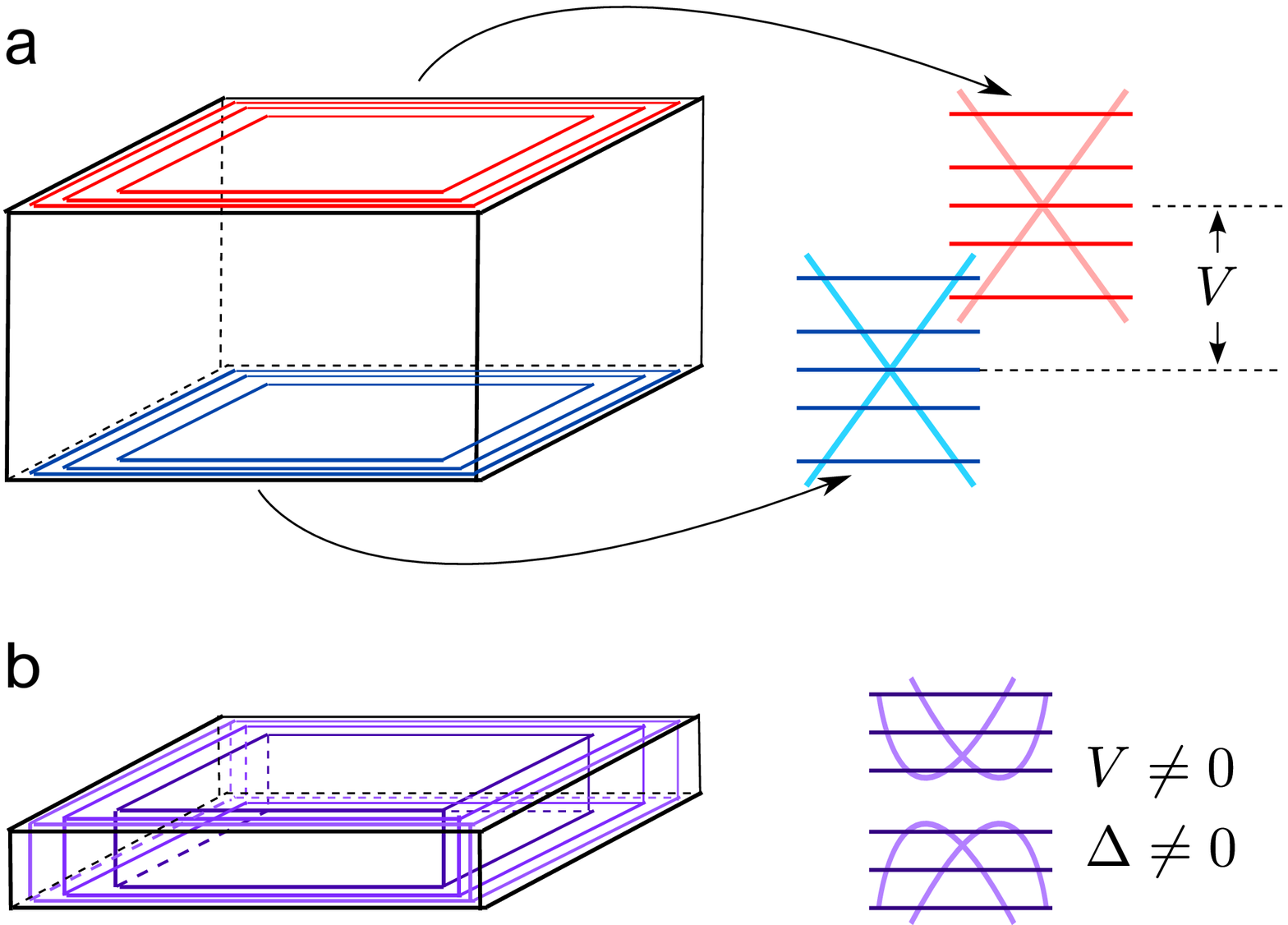} \includegraphics[width=0.55\textwidth]{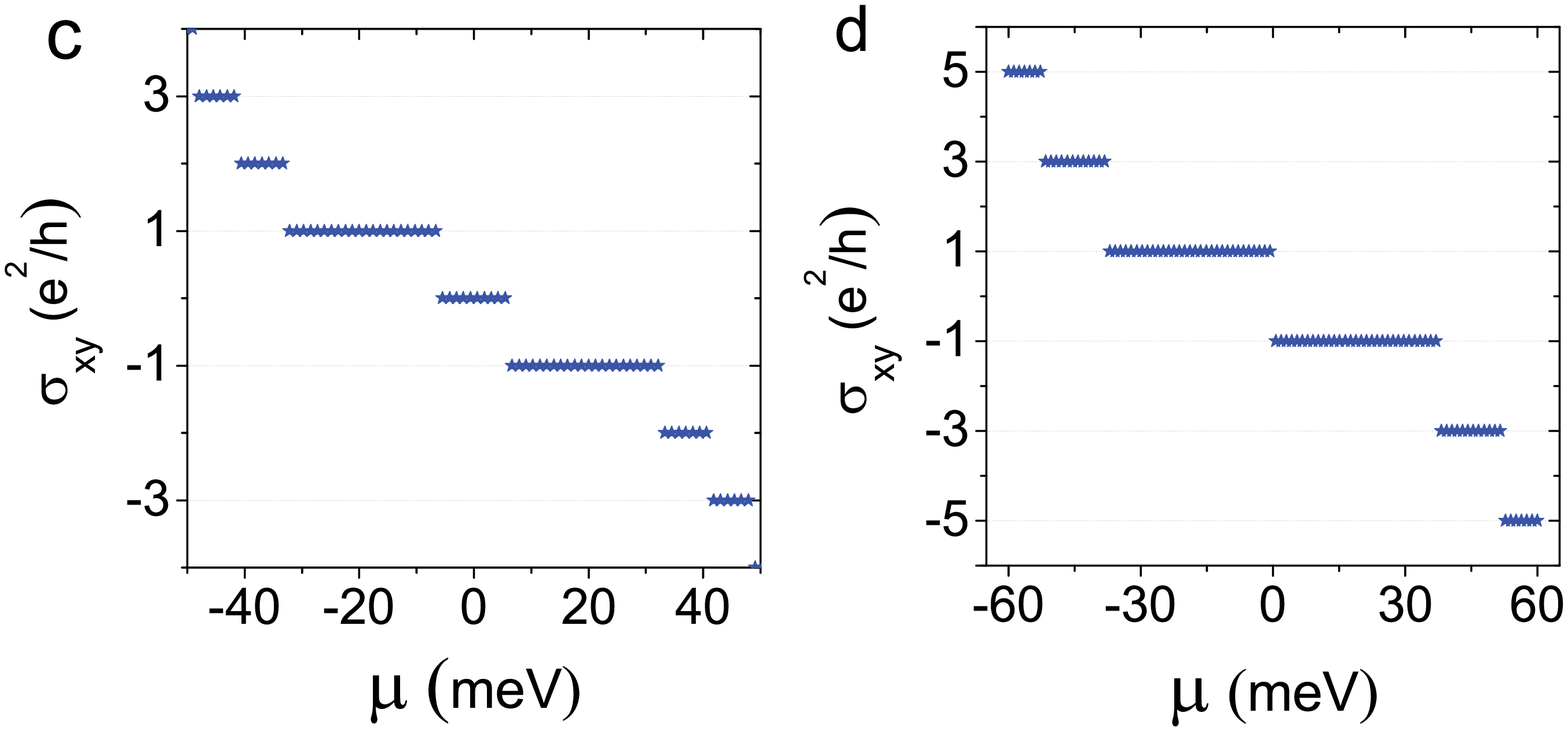}
\protect\caption{\textbf{Two distinct quantum Hall conductance patterns.} (a) Schematics
of LLs of the surface states in a thick topological insulator film
under a perpendicular magnetic field. Each of the top and bottom surfaces
hosts an independent set of LLs. $2V$ is the energy offset between
the top and bottom surfaces, usually induced by SIA and tunable with
a gate voltage. (b) The same as (a) except for a thin film, where
the LLs reside in the whole film rather than at specific surfaces.
$\Delta$ is the finite-size gap opened by the hybridization of the
top and bottom surface states. (c) The pattern of integer quantum
Hall conductance plateaux as a function of the chemical potential
$\mu$ in a fixed magnetic field $\mu_{0}H$ for a general case. (d)
The pattern of odd integer quantum Hall plateaux in the case where
the two sets of LLs of the top and bottom surface states are degenerate. }

\label{Fig1:2Patterns}
\end{figure}

\begin{figure}[tbph]
\centering \includegraphics[width=0.9\textwidth]{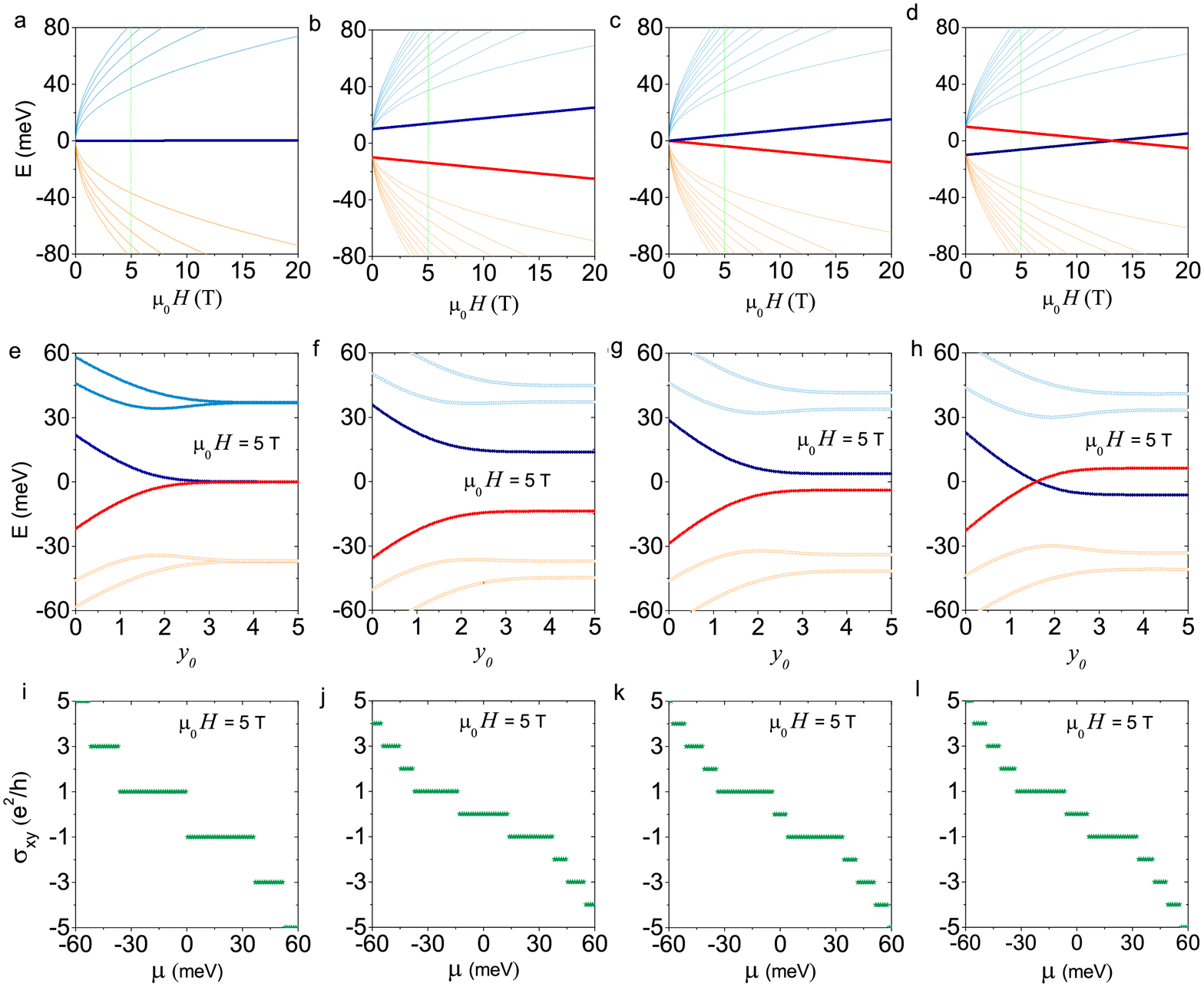} \protect\caption{\textbf{Landau levels and edge states in the absence of SIA.} The
four columns compare cases with different $\Delta$ and $B$, two
parameters in the mass term of the model. From left to right, (i)
$\Delta=0$ and $B\rightarrow0$; (ii) $\Delta B<0$; (iii) $\Delta=0$
and $B\protect\neq0$; (iv) $\Delta B>0$. The top row is for the
fan diagrams, i.e., the energies of LLs as functions of the magnetic
field $\mu_{0}H$. The two LLs of $n=0$ are highlighted. The middle
row is for the energy dispersions of LLs at $\mu_{0}H=5$ T near an
open edge at $y=0$. $y_{0}$ is the position of guiding center in
units of magnetic length $\ell_{B}$. The bottom row is for the Hall
conductance $\sigma_{xy}$ as a function of the chemical potential
$\mu$ at $\mu_{0}H=5$ T. For $\Delta B>0$, the two LLs of $n=0$
cross at a critical magnetic field in (d), showing a field-induced
quantum phase transition from the quantum spin Hall phase in weak
field to trivial band insulator phase in strong field; correspondingly
in (h), the higher hole-like LL and the lower electron-like LL of
$n=0$ cross when approaching the edge, contributing to two conducting
channels with opposite velocities when the chemical potential crosses
them. This characterizes the quantum spin Hall phase with no charge
Hall conductance but a finite quantized spin Hall conductance. For
both cases (ii) and (iii), all electron-like LLs are above all hole-like
LLs, therefore there is no quantum spin Hall phase; For case (i) of
$\Delta=0$ and $B\rightarrow0$, all LLs in (a) are two-fold degenerate
in the bulk, but the degeneracy is lifted in (e) when approaching
the edge. In all cases $B=-500$ meVnm$^{2}$ and $A=300$ meVnm for
comparison, while $\Delta=0$ in cases (i) and (iii), $20$ meV in
case (ii), and $-20$ meV in case (iv). }

\label{Fig2:LL-noSIA}
\end{figure}

\begin{figure}[tbph]
\centering \includegraphics[width=0.9\textwidth]{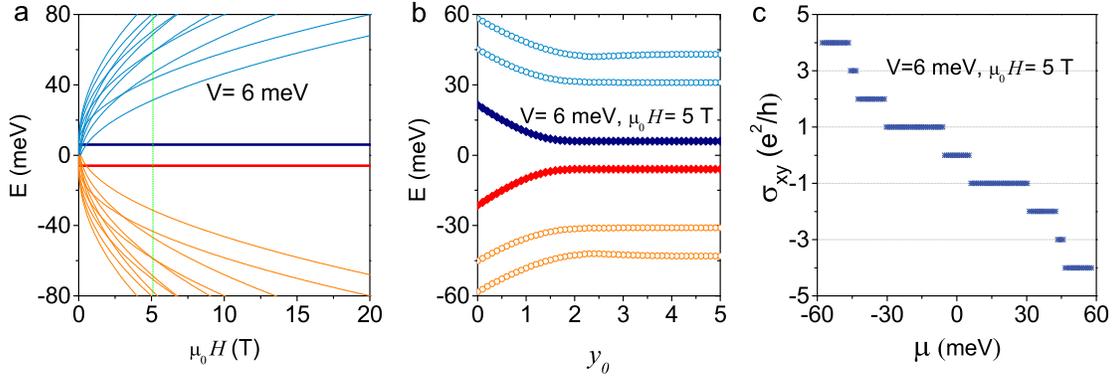} \protect\caption{\textbf{Integer quantized Hall plateaux due to SIA}. For case (i)
with $\Delta=0$ and $B\rightarrow0$ but a finite $V=6$ meV, (a)
the fan diagram, (b) the energies of the two LLs of $n=0$ near the
edge, and (c) the Hall conductance as a function of the chemical potential
$\mu$. SIA breaks the degeneracies of all LLs in the thick film where
$\Delta=0$ and $B\rightarrow0$. As a result, even integer Hall conductance
plateaux also appear. }

\label{Fig3:LL-SIA}
\end{figure}

\begin{figure}[tbph]
\centering \includegraphics[width=0.8\textwidth]{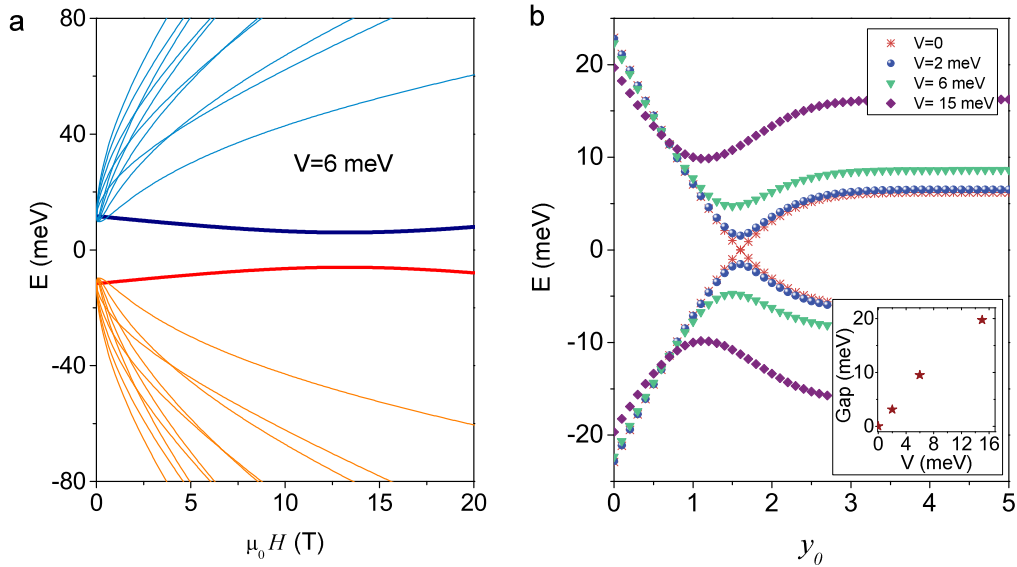} \protect\caption{\textbf{SIA-induced breakdown of the quantum spin Hall phase.} (a)
The fan diagram in the presence of SIA, i.e., $V\protect\neq0$. SIA
turns the crossing between the two LLs of $n=0$ in Fig. \ref{Fig2:LL-noSIA}d
into an anti-crossing. (b) The energies of the two LLs of $n=0$ at
5 T near the edge for different $V$. In the presence of SIA, the
two LLs do not cross near the edge and open an energy gap. Inset:
the gap opened between the two LLs of $n=0$ as a function of $V$.
The parameters are $A=300$ meVnm, $\Delta=-20$ meV, and $B=-500$
meVnm$^{2}$. }

\label{Fig4:BreakDown}
\end{figure}

\begin{figure}[tbph]
\centering \includegraphics[width=0.9\textwidth]{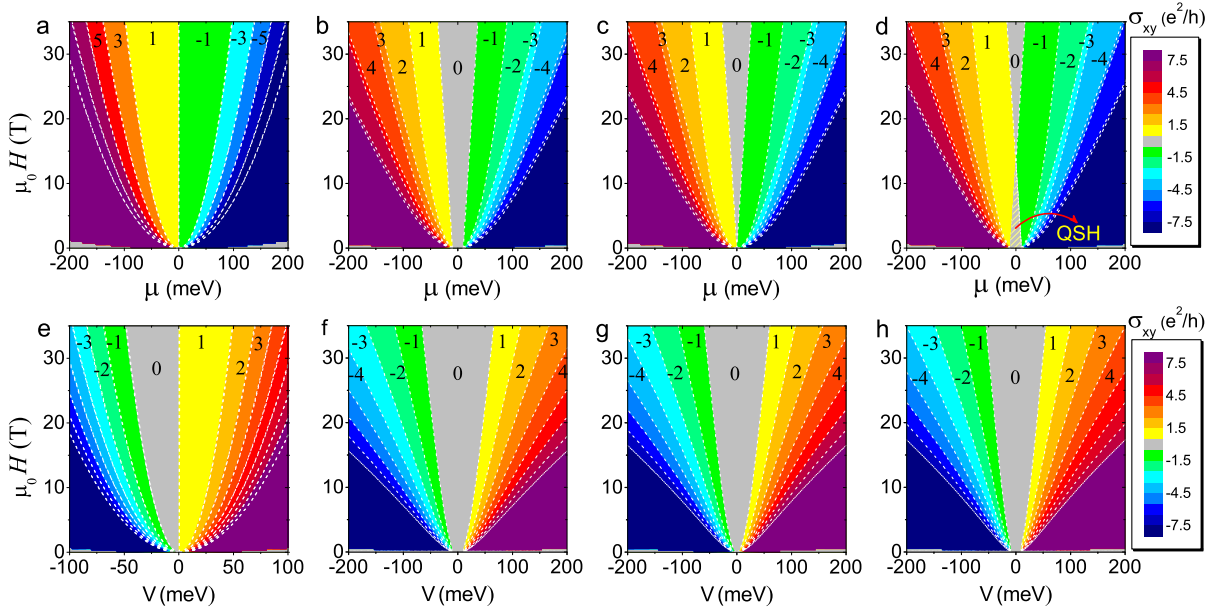} \protect\caption{\textbf{Phase diagrams of the quantum Hall effect in topological insulator
films.} (a-d) The phase diagrams as functions of the chemical potential
$\mu$ and magnetic field $\mu_{0}H$ in the absence of SIA , i.e.,
$V=0$. Different phases are denoted by corresponding Hall conductance
$\sigma_{xy}$ in units of $e^{2}/h$. The white dotted lines are
the boundaries between different phases. The four columns compare
cases with different finite size gap $\Delta$ and $B$. From left
to right, (i) $\Delta=0$ and $B\rightarrow0$; (ii) $\Delta B<0$;
(iii) $\Delta=0$ and $B\protect\neq0$; (iv) $\Delta B>0$. The parameters
for different cases are the same as those in Fig. \ref{Fig2:LL-noSIA}.
$\sigma_{xy}$ is antisymmetric with respect to $\mu$. In (a), there
are only odd integer quantum Hall phases. In (b-d), there are both
odd and even integer quantum Hall phases. In (d), the quantum spin
Hall (QSH) phase is marked. (e-h) The phase diagrams as functions
of $V$ and $\mu_{0}H$ while fixing the chemical potential $\mu=-V-0^{+}$.
Both odd and even integer quantum Hall phases can be induced by changing
$V$ in all the four cases. }

\label{Fig5:PhaseDiagrams}
\end{figure}

\end{widetext}
\end{document}